\documentclass[12pt]{iopart}
\usepackage[dvips]{graphicx}
\begin{document}
%
%

\title{On directed interacting animals and directed percolation}

\author{Milan Kne\v zevi\'c\footnote{E-mail: knez@ff.bg.ac.yu}  and  Jean
Vannimenus\footnote{E-mail: Jean.Vannimenus@lps.ens.fr}}

\address{\ddag Faculty of Physics, University of Belgrade \\
P.O.Box 368, 11000 Belgrade, Yugoslavia
 \vspace{5 mm} \\
\S Laboratoire de Physique Statistique de
l'ENS\footnote{Laboratoire associ\'e au CNRS et aux Universit\'es
Paris~VI et Paris~VII.}
\\24 rue Lhomond, 75005 Paris, France}

\begin{abstract}

 We study the phase diagram of fully directed lattice animals
with nearest-neighbour interactions on the square lattice.
This model comprises  several interesting ensembles
(directed site and bond trees, bond animals,
 strongly embeddable animals)  as special cases  and
its collapse transition is equivalent to a directed bond percolation
threshold.
 Precise estimates  for the animal size  exponents in the different phases
and for the critical fugacities of these special ensembles
are obtained from a phenomenological renormalization group analysis
of the correlation lengths
 for strips of width up to $n=17$.
The crossover region     
in the vicinity of the collapse transition
is analyzed in detail and  the crossover exponent  $\phi$
is determined directly from the singular part of the free energy.
 We show using scaling arguments and an exact relation due to Dhar
 that  $\phi$ is equal to the Fisher    
exponent $\sigma$
governing the size distribution of large directed percolation clusters.

\medbreak



\medbreak


\end{abstract}
%
%
\maketitle

\section{Introduction}

\label{sec:Intro}

 Ever since the seminal work of Broadbent and Hammersley
in 1957 \cite{Hammersley},
the lasting interest  in directed percolation and related statistical models
has been nurtured  by  several motivations.
 Firstly, very anisotropic fractal structures are observed
in Nature in a variety of situations where a bias   
 is introduced by an external field or a gradient,
ranging from river basins~\cite{rivers}
 to stress cracks in metals~\cite{cracks} and to lightning discharges.
 The preferred  direction may alternatively correspond to time, with
fractal patterns appearing in the dynamics of the
systems considered.
  Pomeau~\cite{Pomeau}  thus proposed
 that in simple models of turbulence the spatio-temporal structure
of the active  regions could be described by directed percolation
(DP in the following),
which was later confirmed by extensive numerical
simulations \cite{Grass91, Bohr, Rousseau}.
More generally the question arises if the large-scale properties
 of the observed structures depend on the details of the dynamical processes
occurring on short scales, or if they are more universal and
can be described in terms of purely statistical models.

 Another motivation is that directed systems may be studied
using renormalization group~(RG) tools, providing a possible
approach to non\-equili\-brium phenomena.
The introduction of a preferred direction is
a relevant perturbation in the RG sense,
it is expected to change the universality class of a phase transition
and  modify the critical exponents in a non-trivial way.
In particular  the DP threshold is in the same
universality class as Reggeon field theory~\cite{Sugar}.
In spite of much work, however,  exact results are known only for
simpler variants and DP itself
remains a challenge for theorists,
being one of the few fundamental problems
 to remain unsolved  in two dimensions
 - see~\cite{Hinrichsen} for  a recent review of this model and its position
among models of non-equilibrium phase transitions.
 As pointed out by Grassberger~\cite{Grass97}
it is surprising in view of the central role played by DP
that no detailed experimental tests of the predictions for its critical
behaviour have been carried out.
A  serious difficulty is that in practice
many effects can obscure the expected behaviour at attainable scales,
so it is of interest to explore the crossover regime
between DP and
other  universality classes.

 Directed lattice animals provide another class of fractal structures
that may be relevant for oriented systems,
and it has  been suggested for example that they might
describe the large-scale geometry of river networks~\cite{Moore}.
The basic animal model   
 consists in enumerating all connected  clusters of N sites
 on a lattice but various extensions may be considered
by introducing
local interactions between  
the cluster sites.
One can in this way obtain different statistical ensembles
ranging from loopless structures (lattice trees) to fully compact ones.
In contrast with the situation for DP several
exact results are known for  directed animals on various lattices
\cite{cardy, breuer, dhar0, ndv, dh1}
 - for recent results and a review of what is known rigorously
  see~\cite{bousq} and references therein.

 A remarkable connection between the two types of systems
has been  found by Dhar \cite{dhar3}, who showed that a model of
interacting directed animals (IDA) can be mapped onto a directed
percolation problem
 if certain relations hold between their respective parameters.
For sufficiently strong attractive interactions
the typical animals correspond to
percolation clusters in the dense
 phase above the percolation threshold.
As a consequence the collapse transition of the IDA model is
expected to be of the same nature as the DP threshold,
in contrast to the  isotropic case
where the percolation threshold has been found to correspond
to a higher-order multi-critical point in the phase diagram
of interacting animals~\cite{Coniglio, Rensburg1}.

 We consider in the present work the statistical properties
of  fully directed lattice animals with nearest-neighbour interactions
on the square lattice,
using a transfer matrix approach  to calculate their
correlation lengths
on infinite strips of finite width.
The results are analysed using
a phenomenological renormalization group method
adapted to directed systems \cite{ndv, Kinzel},
which allows
an accurate determination of the critical fugacity.

 The paper is organized as follows:
 in the next section 
 the IDA model and the main quantities of interest are introduced
and we present the correspondences with a percolation problem
and with the generating functions of various simple restricted ensembles.
In section~\ref{numerical} we
apply the phenomenological renormalization (PR) group approach
to non-interacting animals
and check its accuracy  by comparing the results
obtained using strips of widths up to $n = 18$
with the exact values or the best available ones.
 Section~\ref{sec:theta} is devoted to the study of the
behaviour at the collapse transition temperature,
we determine the corresponding critical geometrical
and thermal exponents and compare them with DP exponents.
The low-temperature regime,
which corresponds to the percolating phase of DP,
is considered in section~\ref{sec:lowT}.
In this regime the transition occurring at
the critical fugacity is first-order
and the grand partition function displays an essential singularity,
but the finite-size PR analysis remains applicable and
yields results in excellent agreement
with  Dhar's prediction for the critical line~\cite{dhar3}.
In section~\ref{sec:special}
we obtain  very precise values for the critical fugacities
of various ensembles,
such as bond animals, site or bond trees
and strongly embeddable animals~\cite{flesia1},
and we compare them with published values when these exist.
In section~\ref{sec:crossover}
particular attention is paid to the crossover region
in the immediate vicinity of the collapse transition.
Though the singularity is very weak
 we are able to obtain a direct estimate of the crossover exponent.
 We present analytic arguments
which show that this crossover exponent is  equal
to the Fisher exponent~$\sigma$ for DP~\cite{Stauffer},
in agreement with our numerical results.

\section{The interacting animal model}

\label{sec:animal}

\subsection{Directed animals}

\begin{figure}
\begin{center}
\includegraphics[width=9cm, angle=0]{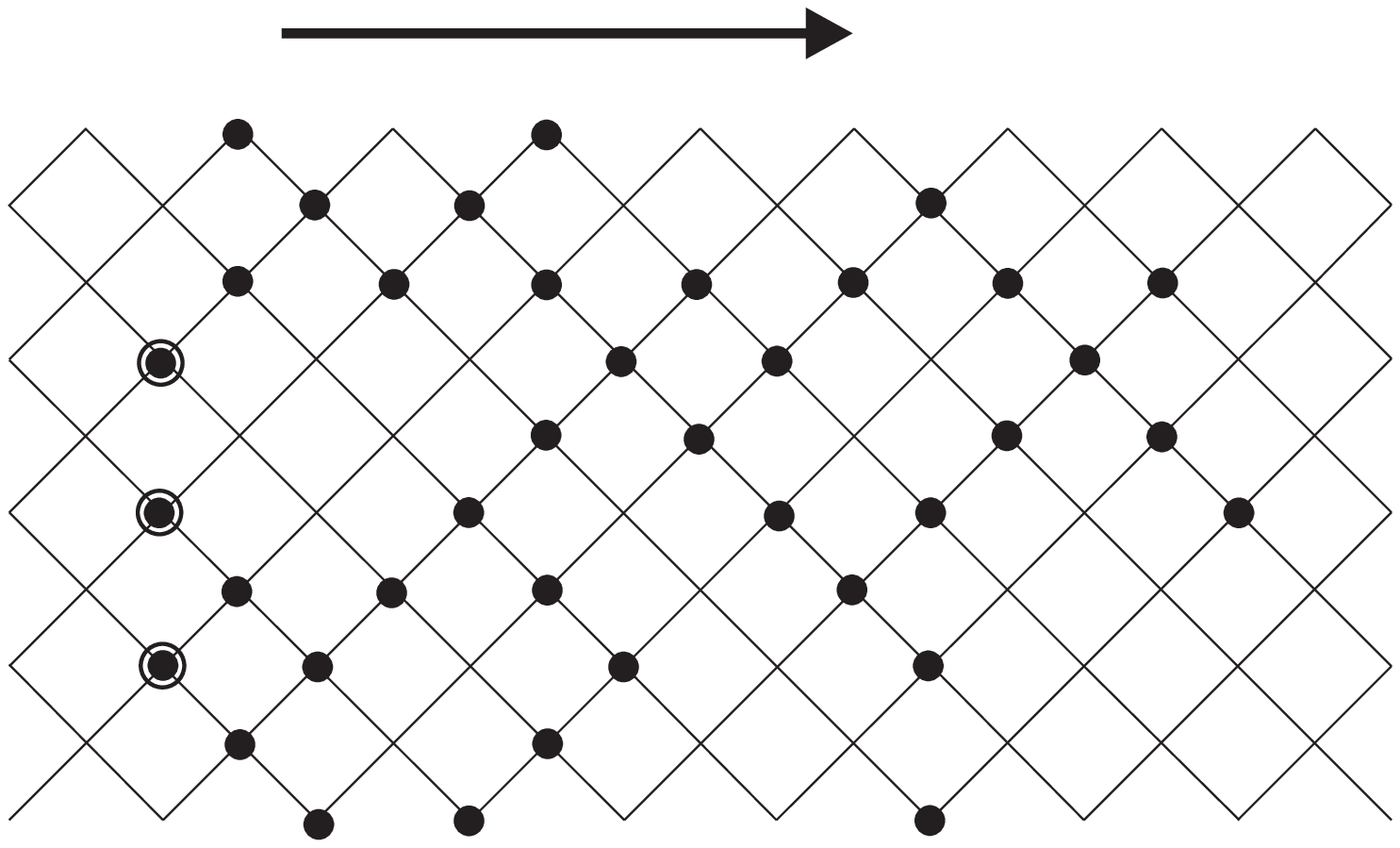}
\end{center}
\caption{ A directed site animal on the square lattice, containing
$N = 33$ sites and $P= 37$ pairs of nearest-neighbour sites
(filled circles), on a strip of width $n= 4$ with periodic
boundary conditions. The root sites are denoted by an extra
circle.}\label{fig:animal}
\end{figure}

A directed animal is a connected cluster of $N$ sites,
 some of which belong to the root, such that any other site
of the animal can be reached from the root by a walk which
never goes opposite to the directed axis (see figure~\ref{fig:animal}).
In the simplest case, non-interacting animals,
 one associates the same weight $x^N$ to each cluster having $N$
sites, so the basic generating function is
 \begin{equation}
 S_0(x) = \sum_{N} \ {\Omega_{N} \: x^N},
\label{eq:F0}
\end{equation}
where $\Omega_N$ is the number of directed connected clusters
 having a single site (the origin) as root.

The problem of enumerating such objects on lattices was
recognized as an interesting combinatorial problem and
received much attention over the last two
decades, in part because it turned out to be possible
to obtain many exact results concerning the $\Omega_N$,
whereas for undirected animals only bounds are known.
In particular the generating function (\ref{eq:F0}) is known
for the square lattice studied here \cite{dhar0, bousq}:
 \begin{equation}
 S_0(x) = \frac{1}{2} ((\frac{1+x}{1-3x})^{1/2} - 1 ).
\label{eq:S0}
\end{equation}

For physical applications the
metric properties of directed animals are perhaps yet more important.
They are best described in terms of two independent characteristic lengths
with asymptotic behaviour of the form

 \begin{equation}
R^{\parallel}\sim c_1 N^{\nu^{\parallel}},
\qquad R^{\perp}\sim c_2 N^{\nu^{\perp}},
\label{eq:xi1}
\end{equation}
where $R^{\parallel}$
provides a measure of the average distance between
the root and the most distant sites along the directed axis, while
$R^{\perp}$
corresponds to the average distance between two extreme
animal sites in the perpendicular (transverse) direction.
 The value of the perpendicular exponent
is known to be $\nu^{\perp}=1/2$ in $d=2$,
from the exact solution for various lattices \cite {dhar0, ndv, dh1}
and in agreement with a field theory argument connecting
the large-scale behavior of directed animals and the dynamics at
the Yang-Lee edge singularity \cite{cardy, breuer}.
The longitudinal  exponent
$\nu^{\parallel}$
has proven much more difficult to obtain and
is known only numerically \cite{dh2,cg}.

\subsection{Branched polymers and interacting animals}

 Interacting animals are often used as simple lattice models
 to describe the behaviour of branched polymers in dilute solutions.
Indeed, field theory calculations \cite{Lubensky} indicate that the
existence of
loops in animals is irrelevant in the RG sense,
so one expects them to belong to the same universality class as branched
polymers
and to share the same large-scale behaviour.
 A simple way to take into account the interactions
with the solvent is to associate an energy   $\epsilon$
with every pair of neighbouring occupied sites.
This energy corresponds to the difference between
the monomer-solvent and the monomer-monomer interaction energies
(one can distinguish between 
chemically bonded  pairs
and others in mere physical contact~\cite{Gaunt},
but the present simple model already contains much of the physics).
At an absolute temperature $T$ the Boltzmann thermal weight
associated with each pair is then
\begin{equation}
w = \exp(\epsilon /k_B T) ,
\label{eq:w}
\end{equation}
where $k_{B}$ is the Boltzmann constant.

 A partition function for polymers made of exactly $N$ monomers
may be defined:
 \begin{equation}
 Z_N(T) = \sum_{P}{\Omega(N, P) \, w^P},
\label{eq:ZN}
\end{equation}
where $\Omega(N, P)$ is the number of clusters of $N$ sites
having $P$  pairs of occupied nearest-neighbors
(note that all possible polymer topologies contribute to $Z_N$,
this corresponds to a model of "living polymers"
where the monomers may rearrange).
At infinite temperature $w=1$ and one recovers the statistics
of an ensemble of equally weighted animals.
A good solvent corresponds to $\epsilon \simeq 0$ ($w \simeq 1$),
a  poor solvent corresponds to $\epsilon >> 0$ (large $w$).
The free energy per site in the thermodynamic limit is given by
\begin{equation}
 f(T) =  \lim_{N \to\infty} -\,\frac{k_{B}T}{N} \log Z_N(T) .
\label{eq:fT}
\end{equation}
It is often convenient to consider the generating function

 \begin{equation}
G(x,w) = \sum_{N}\;{Z_N(w)\,x^N} = \sum_{NP}\;{\Omega(N, P)\,x^N w^P},
\label{eq:G2}
\end{equation}
 which is analogous to a grand partition function
and has a singularity for $x = x_c(w)$, the critical fugacity.
From (\ref{eq:ZN}), (\ref{eq:fT}) and (\ref{eq:G2}) one has
 \begin{equation}
f(T) =  k_{B}T \log x_c(w) .
\label{eq:xc}
\end{equation}

\subsection{Collapse transition and percolation threshold}

\label{sec:perco}

When the temperature is lowered  compact configurations
become more probable and the polymers are expected to shrink
 from an extended state to a  dense  globular state.
 In the limit
$N \to \infty $
a collapse transition occurs between these two states at a well-defined
temperature (for a statistically typical polymer).
This transition is described by a critical point,
usually referred to as the $\theta$-point,
which is analogous to a tricritical point for magnetic systems.
It is believed that the average linear size of the polymers at the
$\theta$-point scales as $N^{\nu_{\theta}}$,
where $\nu_{\theta}$
differs from the  exponents  $\nu$  and  $\nu_{c}$
corresponding respectively to the extended
and dense  states.
Various lattice models have been proposed for the collapse of
isotropic branched polymers~\cite{Gaunt}.
No complete solution of these models on regular lattices is known,
and most of them have been studied
by numerical methods (see, e.g.,~\cite{whitt}) such as
 Monte Carlo simulation, exact enumeration on lattices, or
transfer-matrix calculations coupled with phenomenological
renormalization (PR)~\cite{dh,sv}.

For directed animals the corresponding transition was first studied
by Monte Carlo simulations~\cite{ld},
within an isotropic scaling approach.
Although this assumption did not properly take into account
the intrinsically anisotropic nature of directed animals,
the  value obtained for the $\theta$ temperature is found to be
 in agreement with more accurate approaches.

A central result, due to Dhar~\cite{dhar3}, and rederived
independently in a recent study of the adsorption of interacting
animals~\cite{Rensburg2},
is the existence of a  correspondence  between
the partition function of directed lattice animals
with interactions between first and second neighbours
 and  a site-bond  percolation problem
on the same directed square lattice.
%
In particular the IDA first-neighbour model studied here corresponds
 to  a simple bond percolation problem:
a given animal $A$ occurs with probability
\begin{equation}
\mathcal{P}(A) = \frac{1+q}{p} \; (\frac{p q^2}{1+q} )^N  \;
(\frac{1+q}{q})^P \label{eqn:P(A)}
\end{equation}
 and using (\ref{eq:ZN})
one obtains the relation
\begin{equation}
Pr(N) = \; \sum_{ | A |  = N}{\mathcal{P}(A) }
 = \; \frac{1+ q}{p} \; (\frac{p q^2}{1+ q})^{N} \;  Z_N(w)
\label{ZNw}
\end{equation}
where the sum bears over all clusters of $N$ sites and
\begin{equation}
p = 1-q  = (w-2)/(w-1) \label{perco}
\end{equation}
is the associated bond  percolation probability.
$Pr(N)$ is the probability that for this value of $p$ a
percolation cluster starting from the origin
 contains precisely  $N$  sites.
Hence in the thermodynamic limit the free energy per site is given
for $w>2$   by

\begin{equation}
  \frac{f(w)}{k_B T} = \log (\frac{p q^2}{1+ q}) -
  \lim_{N \to \infty} (1/N) \log Pr (N),
\label{freesite}
\end{equation}

 For isotropic percolation in $d$ dimensions it has been
 proved~\cite{Souillard}  that
above the percolation threshold the decay of $Pr(N)$
is slower than exponential
and that there exists a value $p_1 > p_c$ such that
 for  $p > p_1$   
\begin{equation}
  \log Pr (N) \sim - C N^{1-1/d}  \quad \mbox{if} \quad N \to \infty,
\label{logP}
\end{equation}
but in fact this behaviour is expected to hold for all $ p > p_c$.
  Physically this just means that most sites of a large finite cluster
lie far from its external boundary and belong to its "bulk",
 whose effective volume is of the order of the geometric one,
while the external boundary that needs to be disconnected
from the infinite cluster is not fractal and contains
of the order of
$N^{(d-1)/d}$ sites~\cite{Souillard}.

In the directed case large  finite percolation clusters
 are quite rare above $p_c$  and difficult to study directly,
%
%
but  a proof has been outlined recently that in this case too
the decay of $Pr(N)$ is slower than exponential~\cite{Rensburg2}
and it is natural to expect  that
 $\, \log Pr (N)$
also follows a  power law decay      
for all $p > p_c$.
The second term on the right-hand side of~(\ref{freesite})
is then negligible in the thermodynamic limit
and one obtains a simple equation for the critical line  
in the low-temperature phase~\cite{dhar3}:
 \begin{equation}
x_c(w) =  \frac{p q^2}{1 + q} = \frac{w-2}{w(w-1)^2}
 \qquad \mbox{if} \quad  w > w(\theta) .
\label{eq:xcw}
\end{equation}
A remarkable consequence is that the free energy per site
is not singular on the low-T side of the transition. 
The excellent agreement of this relation
with  our numerical results presented below    
confirms its validity.
As a further consequence 
the critical temperature  of the
IDA model studied here
is related to the critical probability $p_c$
for the directed bond percolation process on the same lattice:
\begin{equation}
w(\theta)=(2-p_c)/(1-p_c).
\label{eq:wtheta}
\end{equation}
The threshold value $p_c$ as well as the values of DP 
critical exponents have recently  been estimated very accurately
using low-density series expansions~\cite{jg,jen1,jen2},
so the thermal critical exponents of collapsing directed animals
are also known with high precision.

Note, however, that the IDA model is more general than DP
as it remains well defined for  repulsive interactions ($w < 1$),
whereas the equivalence with DP is only valid for $w > 2$.
Also, the whole $(x,w)$ plane may be given a physical meaning
if the region $x > x_c(w)$ is interpreted as a "weak gel"
stabilized through application of an external pressure~\cite{dh}.

\subsection{Special ensembles}

\label{special}

Let us remark at this stage that the model defined by (\ref{eq:G2})
contains in its phase diagram several special points
related to
ensembles defined by simple geometrical properties.

Note firstly that for
directed animals the number   $\mathcal{L}$
of independent loops is  just equal
 to the number $N_2$ of sites connected to two occupied predecessors.
Every such site contributes two pairs of occupied neighbours
in~(\ref{eq:G2})
while the other occupied sites  contribute only one, except for the origin
(for simplicity we consider animals rooted at a single site),
so the total number of pairs is $P = 2 N_2 + (N-1 - N_2)$.
The number of loops is then
\begin{equation}
  \mathcal{L} = P - N + 1
\label{eq:Loop}
\end{equation}
and the loop  generating function~\cite{bousq} is given by
\begin{eqnarray}
G_l(y,v) & = & \sum_{N\mathcal{L}}{\Omega(N,\mathcal{L})\,y^N v^\mathcal{L}}
\label{eq:GNl}\\
         & = & v\,G(y/v, v) .
\label{eq:Gl}
\end{eqnarray}
In particular one obtains
the generating function $G_{st}(y)$ of \emph{site trees} (loopless animals):
\begin{equation}
G_{st}(y) = G_l(y,0) = \lim_{w \to 0, x \to \infty} w \, G(x,w)
\label{Gstrees}
\end{equation}
%
with $y = w \, x$ fixed.

 A simple correspondence also exists between non-interacting
directed \emph{bond animals}
(i.e. clusters of connected directed bonds)  and site animals:
For every occupied site having only one predecessor
 consider the bond linking them,
while for the $N_2$ sites with two predecessors consider one of three
possibilities,
either chosing one of the connecting bonds or keeping both of them.
Since $N_2$ is equal to the number $\mathcal{L}$ of loops this correspondence
implies that the generating function  of these bond animals is given by
\begin{eqnarray}
G_b(y) & = & \sum_{N\mathcal{L}}{\, \Omega(N,\mathcal{L})\,y^{N-1-\mathcal{L}}
    \, (2 y + y^2)^\mathcal{L}}
\label{eq:GbN} \\
  & = & \frac{1}{y}\,G_l(y, 2+y)
\label{eq:Gbl} \\
       & = & \frac{1}{x}\,G(x,\frac{2}{1-x}),
\label{eq:Gb}
\end{eqnarray}
with $ x = y/(2+y)$ and where we have used (\ref{eq:GNl}) and (\ref{eq:Gl})
to obtain
(\ref{eq:Gbl}) and (\ref{eq:Gb}) respectively.

 If one now discards from the previous bond configurations
all the loop-forming ones
(i.e. those with two connecting  bonds going to the same site)
one is left with directed \emph{bond trees}.
The correspondence with site animals just discussed
shows that their generating function is  given by
\begin{eqnarray}
G_{bt}(y) & = & \frac{1}{y}\,G_l(y, 2)  \\
       & = & \frac{1}{x}\,G(x,2),
\label{eq:Gbt}
\end{eqnarray}
with $ x = y/2$.

Finally,
\emph{strongly embeddable} animals \cite{flesia1}
 correspond to maximally connected bond configurations,
in the sense that for a given site animal one systematically keeps
the two bonds going to every site with two predecessors
(bond trees correspond to minimally connected configurations).
Their generating function is then simply
\begin{eqnarray}
G_{emb}(y)  & = &
\sum_{N\mathcal{L}}{\, \Omega(N,\mathcal{L})\,y^{N-1-\mathcal{L}}
    \,(y^2)^\mathcal{L}}  \\
 & = & \frac{1}{y}\,G_l(y, y) \\
 & = & G(1,y).
\label{eq:Gemb}
\end{eqnarray}
%
 Detailed numerical results for these various ensembles are presented
 in  section~\ref{sec:special}.

\section{Numerical methods and phenomenological renormalization analysis}
\label{numerical}

\label{sec:PRmethod}

\subsection{Transfer matrix method}

Here we shall use the method of transfer matrices
coupled to a phenomenological renormalization analysis
\cite{barb,night},
that was applied to lattice animals in~\cite{ndv,dh}.
It is convenient to define the
generating function   $G_{0\mathbf{R}}(x,w)$   for the total
number  $\omega_{0\mathbf{R}}(N,P)$  of directed animals
of $N$ sites with $P$ pairs of nearest-neighbors connecting
the points $0$ and $\mathbf{R}$
 \begin{equation}
G_{0\mathbf{R}}(x,w)=\sum_{NP}{\omega_{0\mathbf{R}}(N,P)\,x^Nw^P}.
\label{eq:GR2}
\end{equation}
On a strip of width $n$ with, say, periodic boundary conditions,
linear recursion relations may be written
between the  $\omega_{0\mathbf{R}}(N,P)$  for two successive
values of the distance $R$ along the strip.
The asymptotic behaviour of   $G_{0\mathbf{R}}(x,w)$
is dominated by the largest eigenvalue  $\lambda_n(x,w)$
of the corresponding transfer matrix   $\mathcal{T}_n $ :
\begin{equation}
G_{0\mathbf{R}}(x,w) \sim [\lambda_n(x,w)]^R =
\exp{\left(-{{R}\over{\xi^{\parallel}_{n}(x,w)}} \right)},
\quad  R\to\infty,
\label{eq:GR3}
\end{equation}
which defines the longitudinal (or parallel) correlation length:
\begin{equation}
\xi^{\parallel}_{n}(x,w) =-1/\log\lambda_{n} .
\end{equation}
This correlation length diverges along the critical line
$x = \widetilde{x}_n(w)$,
 where  $\widetilde{x}_n$ is the smallest positive root of
\begin{equation}
\lambda_{n}(\widetilde{x}_n(w),w)=1 \;
\label{eq:lambda4}
\end{equation}
and the generating function (\ref{eq:GR2}) is infinite for
$x > \widetilde{x}_n(w)$.
Let us recall that this line  determines the free energy per site
of a very large animal of width~$n$,
$f_n=k_{B}T\log\widetilde{x}_n(w)$.
In a similar way $\lambda_n(x,w)$ determines the
average site density $\rho_n(w)$ of an animal on the strip~\cite{ndv,dh}
through
\begin{equation}
\rho_n(w)= \left. {{1}\over{n}} \; {{\partial\log[\lambda_n(x,w)]}
\over{\partial \log x}} \; \right| _{x= \widetilde{x}_n} ,
\label{eq:rho5}
\end{equation}
where the derivative is calculated at  $\widetilde{x}_{n}(w)$.

The gel phase, for $x > \widetilde{x}_n(w)$,
will not be studied in the present work,
let us just note that expression~(\ref{eq:rho5}) remains valid
and gives the gel density~\cite{dh},
but the expression of the correlation length is different
(see~\cite{Foster} for a lucid discussion of that point).

\subsection{Phenomenological renormalization in the high - temperature limit}

In the high-temperature limit ($w = 1$) one recovers
non-interacting lattice animals which as mentioned above have been
studied by a variety of methods (see~\cite{ndv,cg,dh} and
references cited therein). Let us recall here just two recent
results. Using a connection of this problem to the continuous-time
dynamics of a one-dimensional lattice gas~\cite{dh2}, Dhar was
able to obtain a very accurate value of the longitudinal size
exponent
  $ \ \nu^{\parallel}=0.817\,33(5)$,
where $(5)$ indicates the estimated
uncertainty for the last digit.
This critical exponent was estimated more recently~\cite{cg}
by the method of series expansions,
which gave $\nu^{\parallel}=0.817\,22(5)$.
Both these results exclude the conjecture
$\nu^{\parallel}=9/11$
which was based on phenomenological renormalization
on strips of width up to  $n=12$~\cite{ndv},
and it is now generally believed that  $\nu^{\parallel}$
is not a rational number.

We shall first extend the latter calculation on strips of width up
to  $n=18$. As we shall show, our estimate of $\nu^{\parallel}$ is
in excellent agreement with the results just quoted, which
provides a useful check of the overall reliability of the present
approach to the case of interacting animals.

According to renormalization-group theory it is expected
that for $n \gg 1$ and  $x-x_c \ll 1$
the longitudinal correlation length satisfies a  finite-size scaling
relation of the form
\begin{equation}
\xi^{\parallel}_{n} \simeq n^{\kappa} F[n^{1/\nu^{\perp}}(x-x_c)],
\label{eq:xi6}
\end{equation}
where the scaling exponent $\kappa = \nu^{\parallel}/\nu^{\perp}$
and the scaling function  $F(z)$
is finite and regular for  $z=0$.
This allows to obtain an estimate  
of the critical fugacity  $x_c$  as the fixed point
 of  phenomenological renormalization equations
involving three different strips.
Using three consecutive widths $(n-1, n, n+1)$
the equations for the fixed point read
\begin{equation}
 \frac{\xi^{\parallel}_{n-1}(x_{c,n})}{(n-1)^{\kappa_n}} =
 \frac{\xi^{\parallel}_{n}(x_{c,n})}{n^{\kappa_n}} =
 \frac{\xi^{\parallel}_{n+1}(x_{c,n})}{(n+1)^{\kappa_n}},
 \label{eq:xifix}
\end{equation}
where $x_{c,n}$ and $\kappa_n$  are respectively estimates
of the critical fugacity and the scaling exponent.
Eliminating $\kappa_n$ in~(\ref{eq:xifix}) gives an equation for $x_{c,n}$:
 \begin{equation}
  \log{\frac{\xi^{\parallel}_{n+1}}{\xi^{\parallel}_{n}}} =
  \quad \frac{\log{\frac{n+1}{n}}}{\log{\frac{n}{n-1}}} \;
  \log{\frac{\xi^{\parallel}_{n}}{\xi^{\parallel}_{n-1}}}
\qquad \mbox{for} \quad x = x_{c,n} \; ,
   \label{eq:xip7}
  \end{equation}
 which would yield the exact value of $x_c$
if the scaling relation~(\ref{eq:xi6}) held rigorously.
 Finite-size estimates $\kappa_n, \nu^{\perp}_n $ and $\nu^{\parallel}_{n}$
of the critical  exponents are then obtained from
\begin{eqnarray}
 \kappa_{n}  & = &  {{\log(\xi^{\parallel}_{n+1}/\xi^{\parallel}_{n})}
  \over{\log[(n+1)/n]}} \; ,
 \label{eq:kappa}  \\
  {{1}\over{\nu^{\perp}_n}} & = & -\kappa_{n} +
  {\log[{(d\xi^{\parallel}_{n+1}/{dx})/(d\xi^{\parallel}_{n}/{dx})]}
  \over{\log[(n+1)/n]}} \; ,
\label{eq:nu88} \\
\nu^{\parallel}_{n} & = &\kappa_{n}\nu^{\perp}_{n} \; ,
 \label{eq:nu8}
\end{eqnarray}
where all ratios and derivatives are evaluated at $x=x_{c,n}$.

To get accurate numerical estimates of the required derivatives
we have used here a perturbative expansion method
for the transfer matrices as described in~\cite{glaus}.
This enabled us to compute
the largest eigenvalue derivatives with an accuracy 
typically better than $10^{-12}$.
We also took advantage of the fact that the transfer matrix
 is singular for $w=1$ (i.e., $det(\mathcal{T}_n) = 0$)
to reduce its size: for instance, for $n=15$ the matrix size could
be reduced from ($1223\times 1223$) to ($198\times 198$).

\begin{table}

\caption{Critical properties of non-interacting directed lattice
animals on the strips of a square lattice with periodic boundary
conditions.
 The finite-size estimates are obtained from the
renormalization group equations (\ref{eq:xip7}-\ref{eq:nu8}).
Extrapolation~1 is based on the three-term 
fit (\ref{eq:nup10}),
while extrapolation~2 relies on the BST algorithm.
Uncertainties in the last quoted digits are due
to the extrapolation procedures and are shown in parentheses. }
\begin{indented}
\item[]\begin{tabular}{@{}llll}
\br $n$ & $\nu^{\perp}_n$& $\nu^{\parallel}_n$ & $x_{c,n}$
\\
\mr

3,4,5   & 0.483452612 &  0.829035839  &  0.333659392915  \\

4,5,6   & 0.489970354 &  0.825744198  &  0.333462353775  \\

5,6,7   & 0.493175785 &  0.823887630  &  0.333395848362  \\

6,7,8   & 0.495010984 &  0.822687170  &  0.333367807660  \\

7,8,9   & 0.496169426 &  0.821844121  &  0.333354128213  \\

8,9,10  & 0.496951869 &  0.821218766  &  0.333346732444  \\

9,10,11 & 0.497507400 &  0.820736423  &  0.333342415719  \\

10,11,12 & 0.497917305 & 0.820353299  &  0.333339741933  \\

11,12,13 & 0.498229180 & 0.820041906  &  0.333338005223 \\

12,13,14 & 0.498472489 & 0.819784080  &  0.333336832461 \\

13,14,15 & 0.498666307 & 0.819567315  &  0.333336014429 \\

14,15,16 & 0.498823451 & 0.819382709  &  0.333335427953 \\

15,16,17 & 0.498952800 & 0.819223752  &  0.333334997468 \\

16,17,18 & 0.499060679 & 0.819085572  &  0.333334674963 \\ \mr

Extrapolation 1 &0.49999(3) &0.81731(5)      &  0.333\,333\,33(1)\\

Extrapolation 2 &0.50000(3) &0.81730(5)      &  0.333\,333\,34(1)\\

Exact & 1/2 & --   &  1/3         \\ \br
\end{tabular}
\end{indented}
\label{tab:highT}
\end{table}

\subsection{Analysis of the numerical results}

In order to extrapolate these finite-size estimates
 two different extrapolation procedures were used.
The first one (to be referred to as  extrapolation~1)
 is based on a simple sequential fit of the form
\begin{equation}
\nu^{\perp}_{n}=\nu^{\perp}+A n^{-\omega_1}+B n^{-\omega_2},
\label{eq:nup10}
\end{equation}
where the estimates of $\nu^{\perp}$ and of $A,B,\omega_1,\omega_2$ can
be obtained from the values of $\nu^{\perp}_{n}$ for five consecutive widths.
One could expect that for large~$n$ these estimates take certain
limiting values, providing the form~(\ref{eq:nup10}) is well chosen. It turns
out, however, that in some extrapolation procedures considered in this
work the effective values of $B$ and $\omega_{2}$ are rather
unstable, which probably indicates that a more complicated form
should be used for the next to leading
correction.
In such cases we prefer to keep only the first two terms on the right
hand side of~(\ref{eq:nup10}), which always seem to lead to a reliable
extrapolation. In either case the new sequences of estimates that we
calculate are monotonic and they depend on~$n$ more weakly than the
original ones.
The final estimates are obtained by plotting the
extrapolated  data versus  $n^{-\widetilde{\omega}}$
were $\widetilde{\omega}$
is adjusted to give a smooth convergence.

The second procedure
used to extrapolate for $n\to\infty$,
which will be referred to here as  extrapolation~2,
is known as the Bulirsch-Stoer (BST) algorithm~\cite{hs}.
This method includes a free parameter~$\omega$ which
has to be adjusted self-consistently to minimize the fluctuations
in the sequences of estimates produced by the BST algorithm. Let
us note here that the ranges of $\omega$ estimated by this
method are in agreement with the corresponding estimates for the
leading correction to scaling exponent
$\omega_1$ from~(\ref{eq:nup10}).

Our finite-size estimates together with the extrapolated values for
$\nu^{\perp}$, $\nu^{\parallel}$ and $x_c$ are presented in table~1.
One can notice the excellent agreement of these extrapolated values
with the exact results $\nu^{\perp}=1/2$, $x_{c}=1/3$, while the present
result $\nu^{\parallel}=0.817\,31(5)$ corroborates the best
earlier estimates of the longitudinal size exponent.

\section{$\theta$-Point behaviour}

\label{sec:theta}

\subsection{Determination of the $\theta$ critical temperature}

It was shown in \cite{ndv} that for non-interacting directed animals
the dependence of the average site density~(\ref{eq:rho5})
on the strip width
is of the form
\begin{equation}
 \rho_{n}(w=1) \sim C \; n^{\frac{1-\nu^{\parallel}}{\nu^{\perp}} -1}
\label{eq:rho11}
\end{equation}
for large $n$ (here and in the following we use $C$ without super-
 or underscript as  generic notation for an undetermined constant).
By analogy with this behaviour and
with the case of isotropic lattice animals~\cite{dh}
this density is expected to have
the following finite-size scaling form in the vicinity of the IDA
$\theta$-point:
\begin{equation}
\rho_{n}(w) \simeq n^{\psi-1}
 H[(w-w(\theta))\,n^{1/\nu_{t}^{\perp}}],  
\label{eq:rho111}
\end{equation}
for $ n\gg 1$ and $ w-w(\theta)\ll 1 $,
with
\begin{equation}
  \psi = (1-\nu_{\theta}^{\parallel})/\nu_{\theta}^{\perp},
\label{eq:psi}
\end{equation}
where $\nu_{\theta}^{\parallel}$ and $\nu_{\theta}^{\perp}$
denote  respectively
the longitudinal and perpendicular size critical exponents at the
 $\theta$-point.
 $H(z)$ is a non-singular scaling function
and $\nu_{t}^{\perp}$, the perpendicular thermal
correlation length exponent, is expected to be the same as
the perpendicular exponent $\nu_{DP}^{\perp}$
for directed percolation. 
Using~(\ref{eq:rho5}) to calculate~$\rho_n$ we have determined the quantities
\begin{equation}
\psi_n(w)={{\log(\rho_{n}/\rho_{n+1})}\over{\log[n/(n+1)]}}+1.
\label{eq:psi12}
\end{equation}
Due to the scaling form~(\ref{eq:rho111}) one expects that
$\psi_n(w(\theta)) \to \psi $ for $n\gg 1$,
so the solutions of the fixed-point equations
\begin{equation}
\psi_{n+1}(w)=\psi_{n}(w),
\label{eq:psi13}
\end{equation}
yield a sequence of estimates
$w_n(\theta)$ and $\psi_{\theta,n}$ for the critical weight  $w(\theta)$ and
 the ratio~$\psi$.

The results are given in table~2.
Note that this two-parameter
 phenomenological renormalization (PR) group method gives a rather
accurate estimate for the Boltzmann critical weight,
$w(\theta)=3.814\,53(1)$.
This estimate is even better than the one
 obtained using relation~(\ref{eq:wtheta})
and the result of a one-parameter PR approach for the directed
 percolation threshold~\cite{Kinzel}.
It is, however, less accurate than the best one:
 $w(\theta)=3.814\,524\,4$,
deduced from (\ref{eq:wtheta}) and the value $p_c~= 0.644\,700\,185(5)$
recently determined using a one-parameter series expansion method
for DP on the square lattice~\cite{jen2}.
So, we shall use this last  value of $w(\theta)$ in order to estimate
 $\nu_{\theta}^{\parallel}$ and $\nu_{\theta}^{\perp}$.

\subsection{Geometrical critical exponents $\nu_{\theta}^{\parallel}$ and
$\nu_{\theta}^{\perp}$ }

There are different methods 
to calculate finite-size estimates of two independent critical exponents
$\nu_{\theta}^{\parallel}$ and $\nu_{\theta}^{\perp}$.
The simplest one is to suppose that
the finite-size scaling form~(\ref{eq:xi6}) holds
for each fixed~$w$, which allows one to apply the approach described
in the preceding subsection. Thus, using three-strip PR for
$w=w(\theta)$ we have obtained two sequences of estimates
 $\nu_{\theta,n}^{\parallel}$ and $\nu_{\theta,n}^{\perp}$.
The results are shown in table~\ref{tab:theta1}.

\begin{table}
\caption{ Finite-size estimates of the critical interaction $w(\theta)$
and of the geometric exponents at the collapse transition.
The estimates
$w_{n}(\theta)$ and $\psi_{\theta,n}$
are obtained from~(\ref{eq:psi13}), while  $\nu^{\perp}_{\theta,n}$
and $\nu^{\parallel}_{\theta,n}$ are computed  using equations
(\ref{eq:xip7}-\ref{eq:nu8}) for $w=w(\theta)=3.814\,524\,4$.
 Extrapolation~1 is based on the sequential fit~(\ref{eq:nup10})
restricted to two terms. }
\begin{indented}
\item[]\begin{tabular}{@{}lllllll}
\br
 $n$ & $w_{n}(\theta)$& $\psi_{\theta,n}$ &
$\nu^{\perp}_{\theta,n}$ &$\nu^{\parallel}_{\theta,n}$\\ \mr 4,5,6
& 3.84492999 & 0.765241160  & 0.434318716  &0.673635137 \\ 5,6,7 &
3.83257414 & 0.760117725  & 0.432326615  &0.675549201 \\ 6,7,8 &
3.82611670 & 0.756970160  & 0.431222068  &0.676643208 \\ 7,8,9 &
3.82242111 & 0.754906943  & 0.430561122  &0.677317531 \\ 8,9,10 &
3.82001545 & 0.753483968  & 0.430142892  &0.677757109 \\ 9,10,11 &
3.81868653 & 0.752461914  & 0.429866869  &0.678056240 \\ 10,11,12&
3.81769328 & 0.751703212  & 0.429678725  &0.678266797 \\ 11,12,13&
3.81699662 & 0.751124448  & 0.429547246  &0.678419088 \\ 12,13,14&
3.81649309 & 0.750672740  & 0.429453585  &0.678531701 \\ 13,14,15&
3.81611976 & 0.750313284  & 0.429385894  &0.678616506 \\ 14,15,16&
3.81583685 & 0.750022425  & 0.429336467  &0.678681344 \\ 15,16,17&
--         & --           & 0.429301202  &0.678731277 \\ \mr
Extrapolation 1 &3.814\,53(1) &0.7479(1) &  0.4293(1) &0.6789(1)\\
\br
\end{tabular}
\end{indented}
\label{tab:theta1}
\end{table}


A more sophisticated analysis \cite{dh} consists in
taking into account explicitly in the scaling function
the two independent relevant fields
that control the tricritical behaviour near the $\theta$-point.
One then writes

\begin{equation}
\xi^{\parallel}_{n} \simeq n^{\kappa_{\theta}} F[n^{1/\nu_1}u,n^{1/\nu_2}v],
\label{eq:xiscaling}
\end{equation}
where to leading order
the scaling fields are related to the model parameters by
\begin{eqnarray}
u &=& a(x-x_{\theta}) + b(w-w(\theta)) + \cdots , \\
v &=& c(x-x_{\theta}) + d(w-w(\theta)) + \cdots ,
\label{eq:uv}
\end{eqnarray}
and
\begin{equation}
\kappa_{\theta} = \nu_{\theta}^{\parallel}/\nu_{\theta}^{\perp} .
\label{eq:kappat}
\end{equation}
The smaller scaling exponent, $\nu_1$, say, is related to the transverse size
of the animals at the $\theta$-point \cite{dh}, so
$\nu_1= \nu_{\theta}^{\perp}$,
while the larger one controls the thermal correlation length, so
$\nu_2= \nu_t^{\perp}$.

 Now fixing $w= w(\theta)$ one can still
use~(\ref{eq:xip7}) and~(\ref{eq:kappa})
to obtain a sequence of estimates $x_{\theta,n}$ and $\kappa_{\theta,n}$
for $x_{\theta}$ and $\kappa_{\theta}$ respectively.
 The extrapolated values for  $x_{\theta}$ (see table~\ref{tab:theta2})
are in excellent agreement with the value deduced from~(\ref{eq:xcw}),
$x_{\theta} = 0.060\,049\,912\,4(10)$.
The main uncertainty comes in fact from the value of $p_c$ used to
estimate $w(\theta)$.

Taking  the partial derivatives with respect to $x$ and $w$
for a given width~$n$
(we follow closely the method and notations of \cite{dh}),
one has at the $\theta$-point
\begin{eqnarray}
\partial_x {\xi^{\parallel}_{n}} &=& n^{\kappa_{\theta}} (a \; n^{1/{\nu_1}}
\partial{F}/\partial{u} + c \; n^{1/{\nu_2}} \partial{F}/\partial{v} ), \\
\partial_w {\xi^{\parallel}_{n}} &=& n^{\kappa_{\theta}} (b \; n^{1/{\nu_1}}
\partial{F}/\partial{u} + d \; n^{1/{\nu_2}} \partial{F}/\partial{v} ).
  \label{eq:partial}
\end{eqnarray}
 The unknown constants $a \; \partial{F}/\partial{u}$ and
$b \; \partial{F}/\partial{u}$ may be eliminated using
another width $m$:
\begin{eqnarray}
 \frac{\partial_x{\xi^{\parallel}_n}}{n^{\kappa_{\theta} +1/{\nu_1}}}
- \frac{\partial_x{\xi^{\parallel}_m}}{m^{\kappa_{\theta} +1/{\nu_1}}} &=&
c \; (n^{1/{\nu_2}-1/{\nu_1}} - m^{1/{\nu_2}-1/{\nu_1}}) \;
\partial{F}/\partial{v} ,\\
 \frac{\partial_w{\xi^{\parallel}_n}}{n^{\kappa_{\theta} +1/{\nu_1}}}
- \frac{\partial_w{\xi^{\parallel}_m}}{m^{\kappa_{\theta} +1/{\nu_1}}} &=&
d \; (n^{1/{\nu_2}-1/{\nu_1}} - m^{1/{\nu_2}-1/{\nu_1}}) \;
\partial{F}/\partial{v} .
 \label{eq:dxinm}
\end{eqnarray}
The 
ratio $c/d$ may in turn be eliminated using a third width $l$,
yielding finally an equation for the exponent
$z = \kappa_{\theta} +1/{\nu_1}$ :
\begin{equation}
\frac{\partial_x{\xi^{\parallel}_n}-(n/m)^{z} \;
\partial_x{\xi^{\parallel}_n}}
{\partial_x{\xi^{\parallel}_n}-(n/l)^{z}\;
 \partial_x{\xi^{\parallel}_l}} =
\frac{\partial_w{\xi^{\parallel}_n}-(n/m)^{z} \;
\partial_w{\xi^{\parallel}_m}}
{\partial_w{\xi^{\parallel}_n}-(n/l)^{z} \;
\partial_w{\xi^{\parallel}_l}} \; .
\label{eq:zeta}
\end{equation}
 The same calculation may be performed again exchanging the terms involving
 $\nu_1$ and $\nu_2$.
One finds that $z' = \kappa_{\theta} +1/{\nu_2} \;$
also satisfies~(\ref{eq:zeta}),
so one expects this equation to have two positive solutions,
from which estimates for $\nu_1$ and $\nu_2$ may be obtained
since $\kappa_{\theta}$ has been estimated independently.
 The numerical results obtained using three consecutive widths
are presented in table~\ref{tab:theta2},
where it is readily seen that the extrapolated value
$\nu^{\perp}_{t} = 1.097(1)$
agrees with the value of the perpendicular percolation exponent
$\nu^{\perp}_{DP} = 1.096\,854(4)$~\cite{jen2}.
The ratio $\kappa_{\theta}$ is also very close to
its directed percolation counterpart
$\nu_{DP}^{\parallel}/\nu_{DP}^{\perp} = 1.580\;745(10)$~\cite{jen2}.

This suggests that the correspondence with critical
directed percolation clusters
is also valid for the geometric properties of IDA clusters
in addition to their thermal critical behaviour.
Indeed, one finds that
$ \nu_{\theta}^{\parallel}/\nu_{DP}^{\parallel} = 0.391\;55(7)$
agrees with the value of the  
exponent
$ \sigma = 1/(\beta_{DP} + \gamma_{DP}) = 0.391\,510(2)$~\cite{jen2}.
This is in direct analogy with the relation
$\nu_c = \sigma_p \, \nu_p$
for isotropic percolation, 
where $\nu_c$ is the  exponent associated
with the typical radius of large clusters~\cite{Stauffer}.
Although interacting animals are related to directed bond percolation
through~(\ref{ZNw}) it was not abvious a priori that
their geometrical exponents should also be related simply
to those of DP, as they depend on the behaviour of
$ G_{0 R} (x,w) $
at fixed $w$ and not along Dhar's line.

\subsection{Density exponents }

Defining  now the thermal density exponent~$\beta_{t}$
through the variation of the density along the critical line :
\begin{equation}
\rho (T) \sim C \; (w - w(\theta))^{\beta_{t}} ,
\label{eq:rhoT}
\end{equation}
one obtains from the scaling relation (\ref{eq:rho111})
and the numerical results above for $\psi$ and $\nu^{\perp}_{t}$
\begin{equation}
  \beta_{t}  = (1- \psi) \; \nu^{\perp}_{t} = 0.2765(4) ,
\label{eq:beta}
\end{equation}
in agreement with the percolation exponent $\beta_{DP} = 0.276\;486(8)$
\cite{jen2}.

 At a fixed temperature, on the other hand, the density
in the gel phase (for $x > x_c(w)$) is expected to vary
close to the critical line as~\cite{Knez2}
\begin{equation}
\rho (x) \sim C \; (x - x_c(w))^{\,\beta_{x}} ,
\label{eq:rhox}
\end{equation}
where the exponent $\beta_x $ depends on the temperature
and is related through scaling relations to the other
critical exponents, yielding
\begin{eqnarray}
\beta_{x} & = & \nu_{\theta}^{\parallel} + \nu_{\theta}^{\perp} - 1
 = 0.1082(1)  \quad \mbox{for} \quad  w = w_{\theta} \; ,
\label{eq:betatheta} \\
& = & \nu^{\parallel} + \nu^{\perp} - 1
 = 0.3173(1) \quad \mbox{for} \quad w < w_{\theta} \;  .
\label{eq:betax}
\end{eqnarray}
We  have not checked these predictions numerically.

\begin{table}
\caption{
Finite-size estimates for $w=w(\theta)=3.814\,524\,4$.
The estimates $x_{\theta,n}$ of the critical fugacity are obtained
using the PR equation~(\ref{eq:xip7}), while
those of the critical exponents
rely on~(\ref{eq:kappa}) and~(\ref{eq:zeta}).
Extrapolation 1 is based on the  sequential fit~(\ref{eq:nup10})
using only two terms (i.e., three parameters),
except for the case of $x_{\theta,n}$,
where a five-parameter fit was used.
As before, the BST algorithm is used in 
 extrapolation procedure 2.}
  \begin{indented}
\item[]\begin{tabular}{@{}lllllll}
\br
 $n$ & $x_{\theta,n}$&
$\nu^{\parallel}_{\theta,n}/\nu^{\perp}_{\theta,n}$
 & $\nu^{\perp}_{\theta,n}$
&$\nu^{\perp}_{t,n}$\\ \mr 6,7,8   & 0.0600496104928 &  1.56917012
& 0.431744944  &1.12856466 \\ 7,8,9   & 0.0600497568952 &
1.57311880  & 0.430976762  &1.12439401 \\ 8,9,10  &
0.0600498269551 &  1.57570908  & 0.430445690  &1.12138438 \\
9,10,11 & 0.0600498630888 &  1.57727568  & 0.430203390 &1.11854334
\\ 10,11,12& 0.0600498828566 &  1.57854083  & 0.429910536
&1.11679217 \\ 11,12,13& 0.0600498941942 & 1.57937827  &
0.429743411  &1.11515184 \\ 12,13,14& 0.0600499009511 &
1.57998468  & 0.429621850  &1.11377199 \\ 13,14,15&
0.0600499051063 &  1.58043100  & 0.429531918 &1.11259417 \\
14,15,16& 0.0600499077280 &  1.58076371  & 0.429464479
&1.11157632 \\ 15,16,17& 0.0600499094171 & 1.58101415  &
0.429413361  &1.11068746 \\
\mr Extrapolation 1&0.0600499124(2)
&1.5805(5)&  0.4294(1) &1.097(1)\\ Extrapolation 2
&0.0600499123(2) & 1.581(1)      & 0.4293(2) &1.097(1)\\ \br
\end{tabular}
\end{indented}
\label{tab:theta2}
\end{table}

\section{Low-temperature region}

\label{sec:lowT}

\subsection{First-order transition and essential singularity}

For $w>w(\theta)$ typical animals are   
in a collapsed state and cover a finite fraction
of the lattice sites in the
thermodynamic limit.
One expects the transition
 at fixed temperature to be first-order~\cite{dh}
and the average site  density   $\rho(x,w)$
 to jump from zero to a finite value on the critical line 
given  by
$x_c = (w-2)/w(w-1)^2$
in that region.

 Using relation (\ref{ZNw})  the animal generating function may be written
\begin{equation}
G(x,w) = \sum_{N}\;{Z_N(w)\,x^N}
= (\frac{w-2}{w}) \quad \sum_{N}\; (x/x_c)^N \; Pr (N).
\label{eq:GPr}
\end{equation}
If we now assume, as discussed  in section~(\ref{sec:perco}), that
for all $p > p_c$ the probability of large finite directed
percolation clusters  varies as
\begin{equation}
\log Pr (N) \sim  - \; C \; N^{\zeta} \; ,
\label{eq:Prasympt}
\end{equation}
 where $\zeta < 1$ ,
equation~(\ref{eq:GPr}) is similar to the one obtained by Fisher~\cite{Fisher}
for the droplet model of condensation,
with $\zeta$ corresponding to a surface energy exponent.
%
He showed that $G(x,w)$ can be expressed as
\begin{equation}
 G(x,w) =   \; \int_0^{\infty} \;\frac{x}{ x_c \; e^t - x} \; f(t) dt \; ,
\label{eq:Gf}
\end{equation}
where the auxiliary function $f(t)$ vanishes at the origin
with an essential singularity of the form
$ f(t) \sim \exp{(-C/t^{\zeta'})} $,
so $G(x,w)$ as well as all its derivatives
remain finite when $x \to x_c$ from below
and $G$ has a very weak singularity of the form
\begin{equation}
\log G_{sing}  \sim  - \; \frac{C}{ (x_c - x)^{\zeta'}} \; ,
\label{eq:Gsing}
\end{equation}
with $\zeta' = \zeta/(1-\zeta)$. In particular $\zeta' = 1$ if
$\zeta$ has the same value $1/2$ as for isotropic $2d$
percolation.

\subsection{Numerical results}

As discussed in \cite{dh, saleur}
the phenomenological renormalization approach may still be used
to determine the critical fugacity,
as the parallel correlation length diverges at the transition.
A direct application of the 3-strip PR approach using
 (\ref{eq:xip7} - \ref{eq:nu8})
yields  results for   $x_c$
that converge very rapidly  to the  value expected from~(\ref{eq:xcw}).
As  can be seen from  table~\ref{tab:lowT} (see the third column),
the finite-size estimates $\widetilde{x}_n$  of~$x_c$ are already
quite good for narrow strips.
To obtain such an accurate numerical check of~(\ref{eq:xcw})
we had to use a mathematical package allowing for
high-precision numbers
(the precision of our input data was set up to 30 digits).

\begin{table}
\caption{Finite-size estimates of the critical fugacity
$\widetilde{x}_n(w=8)$, defined by
$\lambda_{max}(\widetilde{x}_n)=1$, and parallel correlation
length on Dhar's line $\xi^{\parallel}_n = -1/\log\lambda_n(x_D)$,
with $x_D(w=8)= 3/196$. }
\begin{indented}
\item[]\begin{tabular}{@{}lllllll}
\br
 $n$ & $\widetilde{x}_n$& $\widetilde{x}_n - x_D$ & $\xi_n$ \\
\mr
4   & 0.015306166798839312837 & 4.43 $10^{-8}$   &
8.8814536150 $10^4$ \\

5   & 0.015306125866922874256 & 3.42 $10^{-9}$   &  9.2199061032 $10^5$ \\

6   & 0.015306122730364804891 & 2.81 $10^{-10}$  &  9.3327985279 $10^6$  \\

7   & 0.015306122473218244718 & 2.42 $10^{-11}$  &  9.2866324597 $10^7$  \\

8   & 0.015306122451137529752 & 2.16 $10^{-12}$  &  9.1271646904 $10^8$ \\

9  &  0.015306122449176587816 & 1.97 $10^{-13}$  &  8.8871999634 $10^9$  \\

10 &  0.015306122448997933046 & 1.83 $10^{-14}$  &  8.5908642935 $10^{10}$  \\

11 &  0.015306122448981326805 & 1.73 $10^{-15}$  &  8.2562064446 $10^{11}$ \\

12 &  0.015306122448979758112 & 1.66 $10^{-16}$  &  7.8968815836 $10^{12}$ \\

13 &  0.015306122448979607947 & 1.61  $10^{-17}$ &  7.5232899054 $10^{13}$ \\

14 &  0.015306122448979593412 & 1.58 $10^{-18}$  &  7.1433739792
$10^{14}$ \\ \mr
 $\infty$ & 0.015306122448979591837... &  0 \\ \br
\end{tabular}
\end{indented}
\label{tab:lowT}
\end{table}

The exponent~$\kappa_n$, however, is found to diverge,
signalling a failure of~(\ref{eq:xi6}) to describe adequately
the finite-size scaling of~$\xi_n^{\parallel}$.
A more detailed numerical analysis of the behaviour of $\xi_n^{\parallel}$,
for an interaction $w=8$ deep in the low-T region,
shows that it increases exponentially with the width~$n$
if one fixes $x$ at the critical value $x_c(w)$ given by~(\ref{eq:xcw}):
\begin{equation}
\xi_n^{\parallel}(x_c) \sim A \exp(K n),
\label{eq:xin-c}
\end{equation}
and that the fugacity  $\widetilde{x}_n$
at which $ \xi_n^{\parallel} $ diverges
(i.e., $\lambda_n(\widetilde{x}_n) = 1$)
converges exponentially towards~$x_c$
(see~table~\ref{tab:lowT}).
 These results suggest the following finite-size form:
\begin{equation}
\xi_n^{\parallel} \simeq \frac{C}{(\widetilde{x}_n-x)^{\lambda}}
  \simeq \frac{C}{(x_c-x + D \,\exp(-K' n))^{\lambda}}  \; ,
\label{xin-scaling}
\end{equation}
where $\lambda = K/K'$.
 Numerically one finds that $K \simeq K' = 2.21(2)$,
so  
we conjecture that $\lambda = 1$.

 The average mass of the clusters
\begin{equation}
<N(x)> \; = \; x \, \frac{\partial \log G}{\partial x}
\label{eq:<N>}
\end{equation}
 is dominated by the regular part of $G(x)$ and remains finite
when $x \to x_c$ from below at fixed temperature, as well as all
the higher moments of the mass distribution, so the mass and
linear dimensions of the typical clusters also remain finite.
This implies that
the characteristic transverse length to be compared to
the strip width in a scaling expression remains finite at the transition,
which is consistent with the observed divergence of
the effective exponent~$\kappa_n$.

In the zero-temperature limit ($w \to \infty $) the critical
fugacity $x_c \to 0$, with $ x_c \, w^2 \to 1$, and the generating
function $G(x,w)$ is dominated by the configurations with the
largest possible number of pairs of neighbouring occupied sites at
fixed~$N$. These correspond to compact animals, having no internal
holes and the maximum number of independent loops, or equivalently
the minimum number of lateral surface sites (i.e., sites with
either the smallest abscissa on a given row or the smallest
ordinate on a given column of the underlying square lattice). One
would expect these animals to be globular rather than  elongated,
but it is not clear to us how to extract from the transfer matrix
results values for the geometrical exponents of the (extremely
rare) large clusters in that regime.

 Let us note that the present compact animals are different from
the "directed compact percolation clusters"
studied by Essam~\cite{Essam},
which just obey the constraint that two diagonally neighbouring sites
are necessarily followed by an occupied site.
These are themselves a subset of another class of compact clusters
studied in the literature,
named  "fully directed compact"  by physicists \cite{bbs, ps}
and more precisely "diagonally convex directed" (DCD)
by mathematicians \cite{bousqDDC, Feretic}.
DCD animals may be characterized as  having the minimum accessible
perimeter for a given length (i.e., unoccupied sites having
an occupied predecessor).
They have been exactly enumerated with respect to various parameters
\cite{bousqDDC, Inui, Feretic2}
and  are expected to be extremely elongated,
with $\nu^{\parallel} = 1$ and $\nu^{\perp} = 0$
(where $\nu^{\perp}$ is associated with the local width
at a given large distance from the origin, rather than to the
overall animal width~\cite{bbs}).

\section{Special points}
\label{sec:special}
\begin{figure}
\begin{center}
\includegraphics[width=9cm,angle=0]{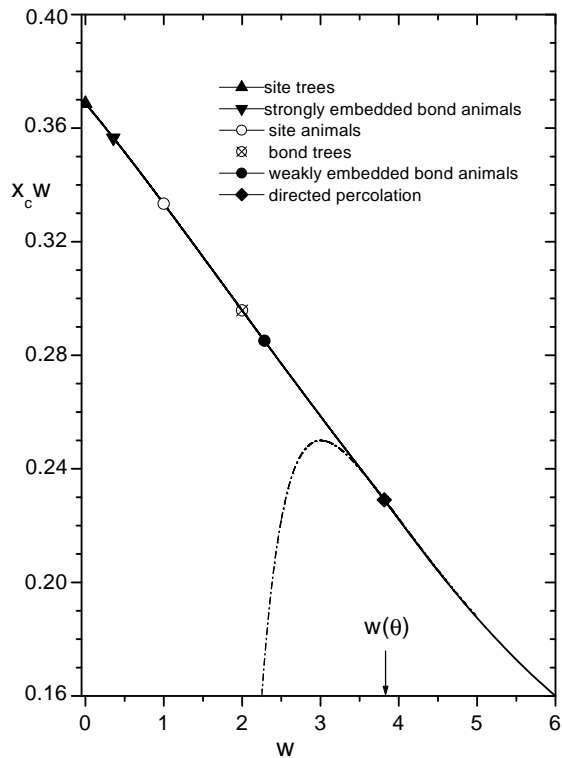}
\end{center}
\caption{Critical line $x_c\, w$ versus interaction parameter~$w$.
The part of the solid curve for $w > w(\theta)$ corresponds to  a
first-order transition and is given by (\ref{eq:xcw}), the
dash-dotted line is its analytic continuation for $w < w(\theta)$.
The symbols denote special values of~$w$ at which $G(x,w)$ is
related to the generating functions of various simple
ensembles.}\label{fig:fugacity}
\end{figure}
As pointed out in section (\ref{special}), the phase diagram of
 interacting directed animals in the~$(w,x)$ plane
contains several special points related to various
ensembles of animals having specific geometric properties.

\begin{itemize}
  \item Thus, from (\ref{eq:Gb}) the generating function of  bond animals
 is obtained on the line $w=\frac{2}{(1-x)}$.
  Using strips of width $(14,15,16)$
we find this line intersects the critical line $x_c(w)$  for
$w_c= 2.285\,08(1)$.
This is in agreement with the value
$ y_b = 0.285\,087\,5(8)$,
obtained for their critical fugacity
by Conway et al.~\cite{cbg}
since in the present notations $w_c= 2 + y_b$.
  \item Bond trees correspond to $w=2$ according to (\ref{eq:Gbt})
and we find that the corresponding $x_c = 0.147\,892\,24(1)$,
in agreement with (but more precise than)
the critical fugacity
$y_{bt} = 2 x_c = 0.295\,785 (10)$ found in~\cite{cbg}.
  \item Site trees correspond to the limit $w \to 0$ with $y = w x$
fixed and finite.
They were studied as Model B in~\cite{ndv} and their critical fugacity
was found there  to be
$y_{st} = 0.368\,649(2)$.
  \item Finally for a site fugacity $x=1$ one recovers the generating
function  of
strongly embeddable animals, see~(\ref{eq:Gemb}),
so their critical fugacity $y_{emb} = w_c(1)$.
We find  $y_{emb} = 0.356\,563\,26(1)$
for this quantity, which does not seem to have been estimated previously
in the literature.
\end{itemize}

A phase diagram summarizing  these various results is displayed
in figure~\ref{fig:fugacity}.
It is noteworthy that for $w < w_{\theta}$ the plotted quantity
$y_c = w \, x_c(w)$,
which according to~(\ref{eq:Gl}) is just
the critical  fugacity for animals enumerated
according to loop number,
varies quasi linearly with~$w$, while this is not at all the case
for the analytic continuation of the
low-temperature expression~(\ref{eq:xcw}). 


\section{Crossover region close to the collapse transition} 

\label{sec:crossover}

\subsection{Critical line and crossover exponent}

 We now focus our attention on the region close to the $\theta$-point.
 As noted above, for $w > w(\theta)$ the critical line $x_c(w)$ is given
by the simple expression~(\ref{eq:xcw})
and a remarkable consequence~\cite{dhar3} is the absence of a singularity
in the bulk free energy per site $f(T)$   
 when $w \to w(\theta)^+$.
The collapse transition is just signalled by a singularity in the
size-dependent  correction term, which may be interpreted as
the vanishing of a surface free energy.
 For $w < w(\theta)$ the critical line departs from the analytic continuation
 of~(\ref{eq:xcw}), which we denote $x_D$    
in the following to avoid confusion.
The singular part is traditionally written under the form
\begin{equation}
x_c - x_D  \simeq C (w(\theta) - w)^{1/\phi},
\label{xsingul}
\end{equation}
where $\phi$ is the crossover exponent.

\begin{figure}
\begin{center}
\includegraphics[width=8cm,angle=0]{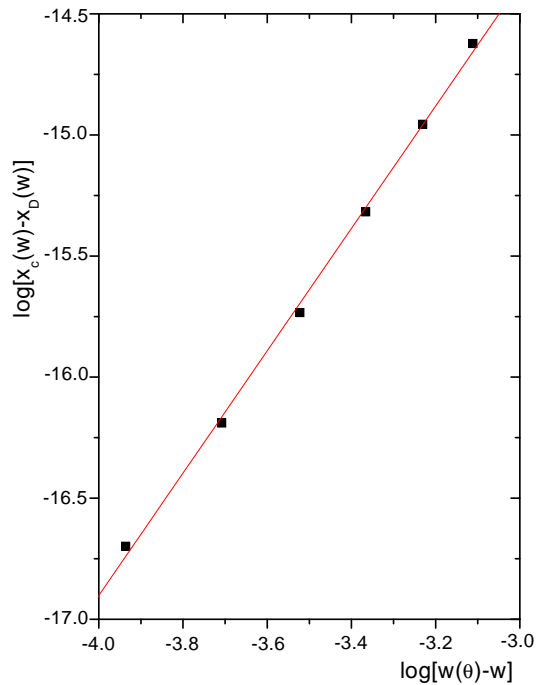}
\end{center}
\caption{Singular part of the critical fugacity, versus distance
to the collapse transition, in a doubly logarithmic plot.
 The line is a linear fit to the  numerical results (filled squares),
whose slope yields the value  $(\phi)^{-1} = 2.53(5)$ for the
crossover exponent. }\label{fig:crossover}
\end{figure}

 In most cases it is  quite difficult to obtain accurate values
for such crossover exponents  from numerical data,
as generally neither the non-singular part of the critical line
nor  the value of the critical temperature
are known with high precision,
making a fit  of the form~(\ref{xsingul}) subject to large uncertainties.
 Here the situation is quite favourable as we can take advantage of the
analytic nature of the free energy in the low-$T$ phase
and of the high-precision knowledge of the critical point $w(\theta)$.
The remaining difficulty comes from the uncertainties due to
the extrapolation procedure
used to obtain $x_c(w)$, which prevent us from analyzing reliably
the data extremely close to $w(\theta)$.
Fortunately there exists an interval of values of~$w$
where the extrapolation uncertainties and
the corrections to~(\ref{xsingul}) are both  small
enough to allow a reliable analysis.

The numerical results are displayed as a log-log plot in
figure~\ref{fig:crossover} and the slope of the curve yields
\begin{eqnarray}
1/\phi & = & 2.53 \pm 0.05 \\
 \phi & = & 0.395 \pm 0.008.
\label{phi}
\end{eqnarray}
  This corresponds to a  weak singularity for the specific heat
measured along the critical line, as the associated
exponent~$\alpha$~\cite{brak} is negative:
\begin{equation}
 \alpha = 2 - 1/\phi = -0.53 \pm 0.05,
\label{alpha}
\end{equation}
 in strong contrast with what is found  for the $\theta$-point
of isotropic animals, where $\alpha \simeq 0.48$~\cite{dh}.
This also implies that in the vicinity of their collapse transition
interacting directed animals
 should display geometric properties close to those
of critical directed percolation clusters up to large scales.

\subsection{Relation with a percolation exponent}

 In order to derive a relation between the crossover exponent $\phi$
and a directed percolation exponent, we rewrite~(\ref{freesite})
under the form
\begin{equation}
  \log x_c = \log {x}_D -
             \lim_{N \to \infty} (1/N) \log Pr (N),
\label{xcxD}
\end{equation}
where ${x}_D = (w-2)/w(w-1)^2$.
%
%
 For $w < w(\theta)$, the bond probability $ p < p_c$
and the cluster probability $Pr(N)$ is expected to decay  exponentially
with $N$,
so the second term on the right-hand side of~(\ref{xcxD})
is no longer negligible when $N \to \infty$.
 Equation (\ref{xcxD}) may then be written
\begin{eqnarray}
  (1/N) \log Pr (N) & \simeq & -\log(x_c/x_D)
           \qquad \mbox{for} \quad  N \gg 1,  \\
 & \simeq & -\frac{x_c - x_D}{x_D}
        \qquad \mbox{for} \quad x_c - x_D \ll x_D, \\
 & \sim &  C [w(\theta) - w]^{1/\phi} ,
\label{PrNw}
\end{eqnarray}
using (\ref{xsingul}) in the last line.
Relation~(\ref{perco}) between $p$ and $w$ still holds in the high-$T$ phase,
so $w(\theta) - w \simeq (p_c - p)/(1-p_c^2)$
and we finally obtain
\begin{equation}
  (1/N) \log Pr (N)  \sim  C (p_c - p)^{1/\phi} .
\label{PrNp}
\end{equation}
Now, close to $p_c$ the probability distribution of cluster sizes is
expected to follow  a scaling law of the form~\cite{Barma}
\begin{equation}
   Pr (N)  \sim  N^{1-\tau} \, \Psi[N^\sigma (p_c - p)],
\label{PrNscal}
\end{equation}
for $N \gg 1$ and $z= N^{\sigma}(p_c-p)$ finite,
where $\Psi(z)$ is a scaling function finite for $z=0$ and
 we have followed the notations of~\cite{Stauffer} for
the Fisher exponents $\tau$ and~$\sigma$.

For  these two expressions for $Pr (N)$ to be compatible
it is necessary that
\begin{equation}
   \phi = \sigma.
\label{phisigma}
\end{equation}
 A very accurate value for~$\sigma$ may be deduced
from Jensen's recent results~\cite{jen2}
and a standard scaling relation between percolation exponents  :
\begin{equation}
    \sigma = 1/(\beta_{DP} + \gamma_{DP}) = 0.391\,510(2),
\label{sigma}
\end{equation}
in good agreement with our numerical value (\ref{phi}) for~$\phi$,
which was obtained without invoking scaling assumptions.


 \section{Conclusion}

We have shown that the combined use of the transfer matrix method and of a
phenomenological renormalization-group analysis gives very accurate results
when applied to interacting directed animals (IDA),
in particular close to their collapse transition.
The present results are still not quite as precise for the tricritical fugacity
and the tricritical exponents as those obtained from the best series
calculations for directed percolation~\cite{jen2},
but the method is more flexible so it could also be applied
without excessive effort
to situations involving more complicated interactions.

\medbreak

\textbf{Acknowledgments}

We thank M. Bousquet-Melou, H. Chat\'e, B. Derrida and D. Dhar
 for useful discussions,
 P. Grassberger for providing several relevant references.
 One of us (M. K) would like to thank the University P.~et M.~Curie (Paris~6)
for financial support and the members of the Laboratoire de Physique
Statistique de l'ENS,
where part of this work has been done, for their kind hospitality.

\medbreak

\end{document}